\documentclass[twoside,twocolumn,9pt]{article}
\usepackage{extsizes}
\usepackage[super,sort&compress,comma]{natbib} 
\usepackage[version=3]{mhchem}
\usepackage[left=1.5cm, right=1.5cm, top=1.785cm, bottom=2.0cm]{geometry}
\usepackage{balance}
\usepackage{times,mathptmx}
\usepackage{sectsty}
\usepackage{graphicx} 
\usepackage{lastpage}
\usepackage[format=plain,justification=justified,singlelinecheck=false,font={stretch=1.125,small,sf},labelfont=bf,labelsep=space]{caption}
\usepackage{float}
\usepackage{fancyhdr}
\usepackage{fnpos}
\usepackage[english]{babel}
\addto{\captionsenglish}{%
  
}
\usepackage{array}
\usepackage{droidsans}
\usepackage{charter}
\usepackage[T1]{fontenc}
\usepackage[usenames,dvipsnames]{xcolor}
\usepackage{setspace}
\usepackage[compact]{titlesec}
\usepackage{hyperref}

\usepackage{epstopdf}

\definecolor{cream}{RGB}{222,217,201}

\begin{document}

\pagestyle{fancy}
\thispagestyle{plain}
\fancypagestyle{plain}{

\fancyhead[C]{\includegraphics[width=18.5cm]{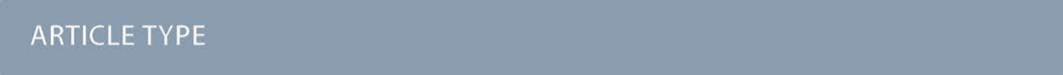}}
\fancyhead[L]{\hspace{0cm}\vspace{1.5cm}\includegraphics[height=30pt]{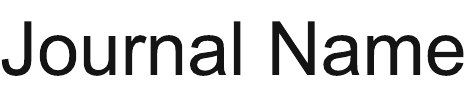}}
\fancyhead[R]{\hspace{0cm}\vspace{1.7cm}\includegraphics[height=55pt]{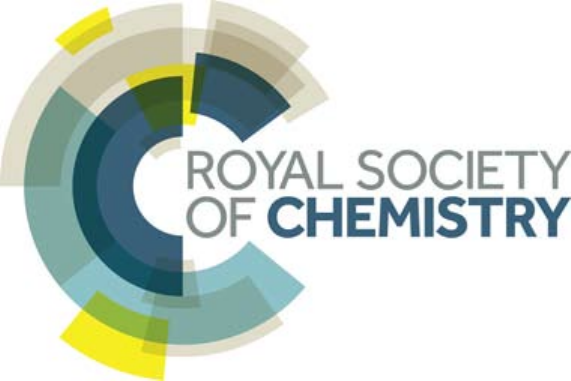}}
\renewcommand{\headrulewidth}{0pt}
}

\makeFNbottom
\makeatletter
\renewcommand\LARGE{\@setfontsize\LARGE{15pt}{17}}
\renewcommand\Large{\@setfontsize\Large{12pt}{14}}
\renewcommand\large{\@setfontsize\large{10pt}{12}}
\renewcommand\footnotesize{\@setfontsize\footnotesize{7pt}{10}}
\makeatother

\renewcommand{\thefootnote}{\fnsymbol{footnote}}
\renewcommand\footnoterule{\vspace*{1pt}%
\color{cream}\hrule width 3.5in height 0.4pt \color{black}\vspace*{5pt}} 
\setcounter{secnumdepth}{5}

\makeatletter 
\renewcommand\@biblabel[1]{#1}            
\renewcommand\@makefntext[1]%
{\noindent\makebox[0pt][r]{\@thefnmark\,}#1}
\makeatother 
\renewcommand{\figurename}{\small{Fig.}~}
\sectionfont{\sffamily\Large}
\subsectionfont{\normalsize}
\subsubsectionfont{\bf}
\setstretch{1.125} 
\setlength{\skip\footins}{0.8cm}
\setlength{\footnotesep}{0.25cm}
\setlength{\jot}{10pt}
\titlespacing*{\section}{0pt}{4pt}{4pt}
\titlespacing*{\subsection}{0pt}{15pt}{1pt}

\fancyfoot{}
\fancyfoot[LO,RE]{\vspace{-7.1pt}\includegraphics[height=9pt]{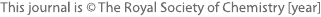}}
\fancyfoot[CO]{\vspace{-7.1pt}\hspace{13.2cm}\includegraphics{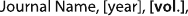}}
\fancyfoot[CE]{\vspace{-7.2pt}\hspace{-14.2cm}\includegraphics{head_foot/RF}}
\fancyfoot[RO]{\footnotesize{\sffamily{1--\pageref{LastPage} ~\textbar  \hspace{2pt}\thepage}}}
\fancyfoot[LE]{\footnotesize{\sffamily{\thepage~\textbar\hspace{3.45cm} 1--\pageref{LastPage}}}}
\fancyhead{}
\renewcommand{\headrulewidth}{0pt} 
\renewcommand{\footrulewidth}{0pt}
\setlength{\arrayrulewidth}{1pt}
\setlength{\columnsep}{6.5mm}
\setlength\bibsep{1pt}

\makeatletter 
\newlength{\figrulesep} 
\setlength{\figrulesep}{0.5\textfloatsep} 

\newcommand{\topfigrule}{\vspace*{-1pt}%
\noindent{\color{cream}\rule[-\figrulesep]{\columnwidth}{1.5pt}} }

\newcommand{\botfigrule}{\vspace*{-2pt}%
\noindent{\color{cream}\rule[\figrulesep]{\columnwidth}{1.5pt}} }

\newcommand{\dblfigrule}{\vspace*{-1pt}%
\noindent{\color{cream}\rule[-\figrulesep]{\textwidth}{1.5pt}} }

\makeatother

\twocolumn[
  \begin{@twocolumnfalse}
\vspace{3cm}
\sffamily
\begin{tabular}{m{4.5cm} p{13.5cm} }

\includegraphics{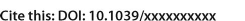} & \noindent\LARGE{\textbf{Effect of halogen dopants on the Li$_{2}$O$_{2}$ properties: Is chloride special?$^\dag$}} \\
\vspace{0.3cm} & \vspace{0.3cm} \\

 & \noindent\large{Henry A. Cortes,\textit{$^{a}$} Ver\'onica L. Vildosola,\textit{$^{ab}$} Mar\'ia Andrea Barral,\textit{$^{ab}$} and Horacio R. Corti \textit{$^{abc}$}} \\

\includegraphics{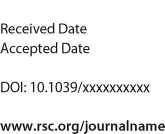} & \noindent\normalsize{There is consensus on the fact that one of the main limitations of the Li air batteries (LABs) is the 
insulating character of \ce{Li2O2} and that it becomes crucial to explore new conduction
paths. Recent studies indicate that doping with chloride increases the ion conductivity
of \ce{Li2O2}, although to a much lesser extent than expected if chloride is assumed
as a donor dopant [Gerbig \textit{et al., Adv. Mater.},2013, \textbf{25}, 3129]. Subsequently, it has been shown that the addition of
lithium chloride, LiCl, to the battery electrolyte increases its discharge capacity while
this effect is not observed with other halogens [Matsuda \textit{et al., J. Phys. Chem. C}, 2016, \textbf{120}, 13360]. This fact was attributed to
an increase in the conductivity of Cl-doped \ce{Li2O2}, but still the
responsible mechanism is not clear. In this work, we have performed first principle calculations
to study the effect of the different halogens (F, Cl, Br, I) as substitutional defects in the electronic and transport properties of
\ce{Li2O2}. We have calculated the formation energies of the different defects and impurities and we analysed how they affect the
activation barriers and diffusion coefficients . 
We have demonstrated that the chloride does not behave like a donor dopant, thus explaining the meager increase
of the ionic conductivity experimentally observed neither it promotes the polaron formation and its mobility.
We have also found that chloride does not present any special behaviour among the halogen series. Our results reveal that all the studied configurations associated with the halogen defects 
	do not derive in metallic states nor extra polarons that would increase considerably the electronic conductivity. This is mainly due to the ionic  characteristics of the \ce{Li2O2} crystal and the capability of the oxygen dimers to adapt its valence rather than to the nature of the dopant itself. }

\end{tabular}

 \end{@twocolumnfalse} \vspace{0.6cm}

  ]

\renewcommand*\rmdefault{bch}\normalfont\upshape
\rmfamily
\section*{}
\vspace{-1cm}


\footnotetext{\textit{$^{a}$Departamento de F\'{i}sica de la Materia Condensada, Comisi\'on Nacional de Energ\'{\i}a At\'omica (CNEA), Buenos Aires, Argentina. E-mail: vildosol@tandar.cnea.gov.ar, hrcorti@tandar.cnea.gov.ar}}
\footnotetext{\textit{$^{b}$Instituto de Nanociencia y Nanotecnolog\'{\i}a (INN CNEA-CONICET), Buenos Aires, Argentina.}}
\footnotetext{\textit{$^{c}$Instituto de Qu\'{\i}mica F\'{\i}sica de los Materiales Medio Ambiente y Energ\'{\i}a (INQUIMAE CONICET), Facultad de Ciencias Exactas y Naturales, Buenos Aires, Argentina}}

\footnotetext{\dag Electronic Supplementary Information (ESI) available: Details about the calculation of the different chemical potentials, the estimation of the the activation barrier of hole polarons and the thermodynamic of the \ce{Li2O2} decomposition reaction. See DOI: 10.1039/b000000x/}



\section{Introduction}

Rechargeable lithium-air batteries (LABs) have recently attracted considerable attention as a possible energy storage device, mainly
for electric vehicles and other energy applications, due to their high theoretical specific energy\cite{Gallagher2014}. However, so far, a practical
implementation has been severely obstructed by a number of technical challenges, including: high charging potential and reduced capacity
at high discharge rates that lead to low discharge-charge cycle efficiency and poor cathode and electrolyte stability, that dramatically limit
cyclability \cite{Luntz-2014,Lu-Gallant-2013,McCloskey-2015}.

In the absence of undesirable side reactions, lithium peroxide (\ce{Li2O2}), insoluble in aprotic organic solvents, is deposited as the primary
discharge product at the porous air cathode of the LAB.
Different morphologies of the deposited \ce{Li2O2} during LAB discharge have been observed, from thin films (5-10 nm in thickness) covering the cathode uniformly, to
large particles or toroids with size in the range of 100-1000 nanometers.
In order to explain the different morphologies, two  mechanism of \ce{Li2O2} formation in the LAB
have been proposed \cite{Johnson2014}, the surface and solution mechanism.
The prevailing one is determined by the competition between the solubility of lithium superoxide (\ce{LiO2}) in the electrolyte and its adsorption on the cathode surface, an intermediate that finally disproportionates, forming solid \ce{Li2O2} and evolving oxygen\cite{TAN2017155}.

Contrarily to lithium oxide (\ce{Li2O}), that is an ionic conductor\cite{li2o}, \ce{Li2O2} is a wide band gap insulator\cite{Ong-2012,Gerbig-2013}. Its huge electrical resistance is considered one of the main limitations
of the Li-air batteries. During discharge the \ce{Li2O2} deposition passivate the cathode surface inhibiting large cell capacity, while during charge its insulating
nature results in large overpotentials that degrades the battery efficiency.

It was soon realized that a detailed understanding of the electron conduction mechanism in \ce{Li2O2} would contribute to improve the performance of LAB.
Thus, several first principles studies were performed in order to elucidate the nature of the charge carriers in \ce{Li2O2} \cite{Hummelshoj-2010,Radin-2012,Ong-2012,Kang-2012}.
Hummelsh\o j \textit{et al}. \cite{Hummelshoj-2010} performed the first DFT calculation of the free energy of intermediates during the discharge and charge of a LAB. They concluded that quite mobile bulk Li vacancies can be created during LAB charge and \ce{Li2O2} becomes a p-type conductor, while during discharge there is a high concentration of Li vacancies at the surface that generates a metallic state that, probably, could sustain high current densities. Similar surface metallic character was reported later by Radin \textit{et al}\cite{Radin-2012}. 
In this line, DFT calculations by Geng \textit{et al}\cite{Geng-2013}. suggested that the electronic conductivity at O-rich grain boundaries of polycristalline \ce{Li2O2} could be enhanced as compared to bulk while
amorphous \ce{Li2O2} has been shown to have similar band gaps as compared to the crystalline phase, but its ionic conductivity is predicted to be 12 orders of magnitude larger ($\sim$2.10$^{-7}$ S.cm$^{-1}$) due to the increase of concentration and mobility of negative lithium vacancies\cite{Tian-2014}.

The possibility of polaronic conduction in peroxide species were also reported\cite{Ong-2012,Kang-2012,DFT+U,Radin-2013}. Ong \textit{et al}.\cite{Ong-2012} showed, using Heyd-Scuseria-Ernzerhof (HSE) screened
hybrid functional, that holes can be self-trapped at the O$_{2}^{-2}$ peroxide ions forming small hole polarons (O$_{2}^-$ superoxide ions), that are
strongly bound to lithium vacancies (V$_{Li}$). Thus, the conductivity would take place mainly via polaron-V$_{Li}$ migration, being the energy barrier for the free vacancy diffusion relatively low (0.4 eV).
At the same time, Kang \textit{et al}.\cite{Kang-2012}, also employing HSE functional, reported that electrons can be localized at the O$_{2}^{-2}$ peroxide ions, occupying its LUMO orbital,
forming small electron polarons (O$_{2}^{-3}$ ions). The migration of the electron polarons occurs via thermally activated hopping, with very high energy barriers leading to extremely
low calculated electron mobility.

Garcia-Lastra \textit{et al}.\cite{DFT+U} revisited the hole and electron polaronic conduction in \ce{Li2O2} using Perdew-Burke-Ernzerhof (PBE) functional with Hubbard correction (PBE+U) showing that the formation of
polarons (both hole and electron) is stabilized with respect to the delocalized states. They concluded that excess electrons are easier to localize than the holes, while the energy barrier for hole hopping are
similar to that reported by Ong \textit{et al}.\cite{Ong-2012}, but the barrier for electron hopping is above 1 eV. By resorting to either the HSE hybrid functional and the many-body perturbation theory (GW),
Radin and Siegel\cite{Radin-2013} concluded that charge transport in \ce{Li2O2} is mediated by both, the migration of V$_{Li}$ and the hopping of hole polarons (p$^+$).

From the experimental point of view, H\o jberg \textit{et al}.\cite{hojberg-2015} have observed that the impedance of a LAB during charge was low compared to the end of discharge,
supporting the model proposed by Luntz \textit{et al}.,\cite{Luntz-2013} where the electronic conductivity is improved upon charge by a reduction of the tunneling barrier due to the
alignment of the \ce{Li2O2} valence band maximum close to the Fermi energy. In that model for ultra-thin films it is assumed that charge transport through \ce{Li2O2} at room temperature, at practical
current densities, is based principally on hole tunneling, while at higher temperatures hole polaron conductivity plays a dominant role. However, Luntz and coworkers\cite{Varley-2014}
observed later that positively charged species, such as oxygen vacancies, and hole polarons, are favorable native vacancies that hinders charge transport by scattering of the tunneling holes.

The first reported impedance spectroscopy and DC conductivity studies of microcrystalline \ce{Li2O2} was performed only recently by Gerbig \textit{et al}.\cite{Gerbig-2013}.
The results demonstrate that ionic lithium defects are the majority carriers with ionic conductivity $\sigma_{ion} \approx$ 10$^{-10}$-10$^{-9}$ S.cm$^{-1}$ at 100 $^{\circ}$C, while the electronic conductivity is around
two orders of magnitude lower at the same temperature and it increases with the oxygen pressure, indicating hole conduction (electrons are consumed upon oxygen incorporation). These results were confirmed later by Dunst \textit{et al}.\cite{Dunst-2014}, who also showed that the overall conductivity of the microcrystalline sample used as reference was very low (3.4x10$^{-13}$ S.cm$^{-1}$), but ball-milling leads to a conductivity increase up to 1.1x10$^{-10}$ S.cm$^{-1}$, being the electronic contribution less than 10\% of the total.

Overall, it can be concluded from the full theoretical and experimental evidence that alternative conduction paths are needed into the \ce{Li2O2} to obtain high capacity and power in LAB at room temperature.
With this aim, some efforts have been recently focused on the effects of dopants on the charge transport in \ce{Li2O2} \cite{Vladimir-2013,Radin-2015}.  Timoshevskii \textit{et al}.\cite{Vladimir-2013} showed by
DFT calculations that doping \ce{Li2O2} with substitutional Si, non-polaronic conducting states would appear in the band gap increasing significantly the electronic mobility due to tunneling
between \ce{SiO6} $\sigma^* $-states. On the other hand, combining first-principles calculations with continuum-scale transport theory, Radin \textit{et al}.\cite{Radin-2015} proposed a multi-scale model
which suggests that thick \ce{Li2O2} deposits doped with traces of Co can support larger recharge current densities with only minimal overpotentials.
The effect would be due to the fact that doping enhances charge transport by shifting the balance of lithium vacancies and hole polarons.

Experimentally, there are few works that studied the role of dopants\cite{Gerbig-2013,Matsuda-2016, Matsuda-2017}.
In an attempt to increase the ionic conductivity, Gerbig \textit{et al}.\cite{Gerbig-2013} doped \ce{Li2O2} with Cl. They assumed that Cl$^{-}$ behaves as a donor dopant
promoting the increase of Li vacancy concentrations that, in turn, would increase $\sigma_{ion}$.  The reported results show that the measured $\sigma_{ion}$ of Cl-doped \ce{Li2O2} has the same order of magnitude as in the undoped sample (increasing only a factor of three). The authors pointed out that the observed increase of $\sigma_{ion}$ was less than expected, since the chemical analysis indicates a
composition of 4900-6000 ppm Cl \textit{vs} 15 ppm of metal cation donor impurities.

Recently, Nakanishi and coworkers\cite{Matsuda-2016} explored the effect of halogen doping in LABs. They showed that when adding LiCl to the electrolyte, the LAB energy capacity is notably larger
than when adding LiF, LiBr, or LiI. The authors found up to 7 mol \% Cl respect to Li atoms was incorporated into the \ce{Li2O2} deposits, and they performed a conductive AFM analysis concluding that the large capacity obtained in the Cl-doped system is connected with an increment of the electronic conductivity.

Even when doping \ce{Li2O2} is a potential route in the search of improving the overall perfomance of LABs, there are still several important open questions, namely: can the dopants effectively improve the bulk conductivity of \ce{Li2O2}?; does dopants affect the morphology of the deposits or the conduction mechanism?; is the process of dopant incorporation from the electrolyte reversible and/or affect the cyclability?

In this work, we perform first-principles calculations to study the electronic properties of halogen doped \ce{Li2O2}. We calculate the formation energies of F, Cl, Br, and I as subtitutional dopants, the energetics regarding electron and hole polaron defects, Li vacancies, and several complexes. We then calculate the activation barriers of the more favorable charged defects and compare the doped and undoped cases.

\section{Computational methodology}

In this work, we study the effect of halogen dopants on the electronic, structural and transport properties of bulk \ce{Li2O2} by means of first
principle calculations based on the Density Functional Theory (DFT)\cite{DFT}, as implemented in the Vienna \textit{ab initio} simulation package (VASP)\cite{Kresse-1996-1,Kresse-1996-2}. 
All the results reported in the main part of this work are obtained  
considering the screened Heyd-Scuseria-Ernzerhof (HSE) hybrid functional \cite{HSE06}, using the PBE functional as the base with which the Hartree-Fock exchange is combined taking $\alpha=0.48$ for the mixing 
parameter and the performing a $\Gamma$-point calculation.
Due to the lack of reported experimental data of the band gap of \ce{Li2O2}, we report the results obtained for this particular value of $\alpha$ since it better reproduces the
band gap calculated with the more accurate self-consistent $GW$\cite{Radin-2013,GW1,GW2}. 

In order to calculate the formation energies of the different defects, we need to simulate isolated impurities so that the supercell should be as large as possible.
We use a 3x3x2 supercell of \ce{Li2O2} with one dopant per cell. We also calculate the formation energies of the most relevant charged defects using a 4x4x2 supercell, and verify that they reasonably converge, in agreement with previously reported results\cite{Radin-2013}. 
We calculate the equilibrium lattice parameters, $a$ and $c$, of the unit cell, obtaining 3.06 \AA\   and 7.48 \AA, respectively, in reasonable agreement with the experimental reported values (3.14 \AA\ and 7.64 \AA, respectively)\cite{wagemaker-2014}. In order to avoid spurious strain in large supercells, we keep the values of $a$ and $c$ fixed for the supercells containing the defects, and we allow the internal positions to relax. Curiously, the structural lattice parameters obtained with PBE agree better with the experiments but, unfortunately, due to large spurious self-interaction of this functional, the band gap is considerably subestimated, and it is not able to stabilize the hole polarons, that, as it will be shown next, is the most favorable positive defect in \ce{Li2O2}.

The formation energies of different halogen dopants in \ce{Li2O2} were calculated before and reported in the supplementary information provided by Radin \textit{et al}\cite{Radin-2015}.
Several types of defects were tested, either substitutional and interstitial ones, with different charge states. In this work, we first reproduce the formation energies reported by Radin \textit{et al}\cite{Radin-2012}, and then we calculate the formation energies of the other types of related defects, and analyse the effect of dopants on the structural, electronic and transport properties. We put special emphasis in assessing the role of halogen ions as donor dopants in \ce{Li2O2}.

As it was previously shown that the interstitial sites are energetically unfavorable\cite{Radin-2015}, we focus on substitutional halogen defects.
Besides, the X-Ray diffraction spectra (XRD) of the discharge product of the LAB reported by Matsuda \textit{et al}. \cite{Matsuda-2016}, revealed the same pattern corresponding to \ce{Li2O2} both in the presence and in the absence of LiCl in the electrolyte. Also the XRD analysis of the Cl-doped synthesized samples in Ref. \citenum{Gerbig-2013} indicates no LiCl formation.
These results ruled out the formation of significant segregation or secondary phases containing Cl in the cited experiments.
Thereby, we simulate bulk Li$_2$X$_x$O$_{2-2x}$ (X=F, Cl, Br, I) considering that the dopants are substitutional defects homogenously distributed along the discharge product.

The possible halogen substitutional sites are: Li-octahedral, Li-trigonal, O, and the O$_2$ dimer. For all the halogens, the more favorable substitutional site is the $O_2$ so that, from now on, we will describe the results obtained out of halogen substitution of the O$_2$ dimer.
Beside the most common intrinsic defects of \ce{Li2O2}, different charged states of the halogen doped system are considered together with several types of complexes,  halogen-vacancy and halogen-polaron as will be shown later on.

The formation energy of a given defect $X$ with charge $q$, E$_f$($X^q$) is calculated according to:
\begin{equation}
        E_f(X^q)=E_0(X^0)-E_0(bulk)-\sum_i n_i\mu_i+q\varepsilon_{F} + E_{MP1},
\label{energy}
\end{equation}

\noindent
where n$_i$ is the number of atoms of the $i^{th}$ species in the defect, $\mu_i$ is the chemical potential of that species, $\varepsilon_F$ is the Fermi level, and E$_{MP1}$=$\frac{q^2\alpha}{2\varepsilon L}$, is the Makov-Payne monopole size correction\cite{PhysRevB.51.4014}, being $\alpha$ the Madelung constant  for a simple cubic lattice of side $L$= 10.25 \AA, and $\varepsilon$= 10 is the average permitivity of \ce{Li2O2} calculated using density functional perturbation theory (with PBE functional).
The Fermi level is determined by choosing the point along the \ce{Li2O2} band gap that satisfies the charge neutrality, $\sum_iq_Xc^0(X^q)=0$, where $c^0(X^q)$ is the corresponding equilibrium concentration for the defect $X^q$. This concentration is related to the formation energy by $c^0(X^q)=D_Xe^{-\frac{E_f(X^q)}{k_BT}}$, being $D_X$ the density of defect sites.
As discussed by Radin \textit{et al}.\cite{Radin-2015},  and the references there in, this condition is suitable for large systems as compared to the screening length of \ce{Li2O2} ($\sim$ 10 nm), as it is the case of the two experimental works reported on Cl-doped \ce{Li2O2}.
In the Supplementary information we give details on the calculation of the
chemical potential of the different species involved.

\section{Results and discussion} 

\begin{figure}
\centering
\includegraphics[height=15cm]{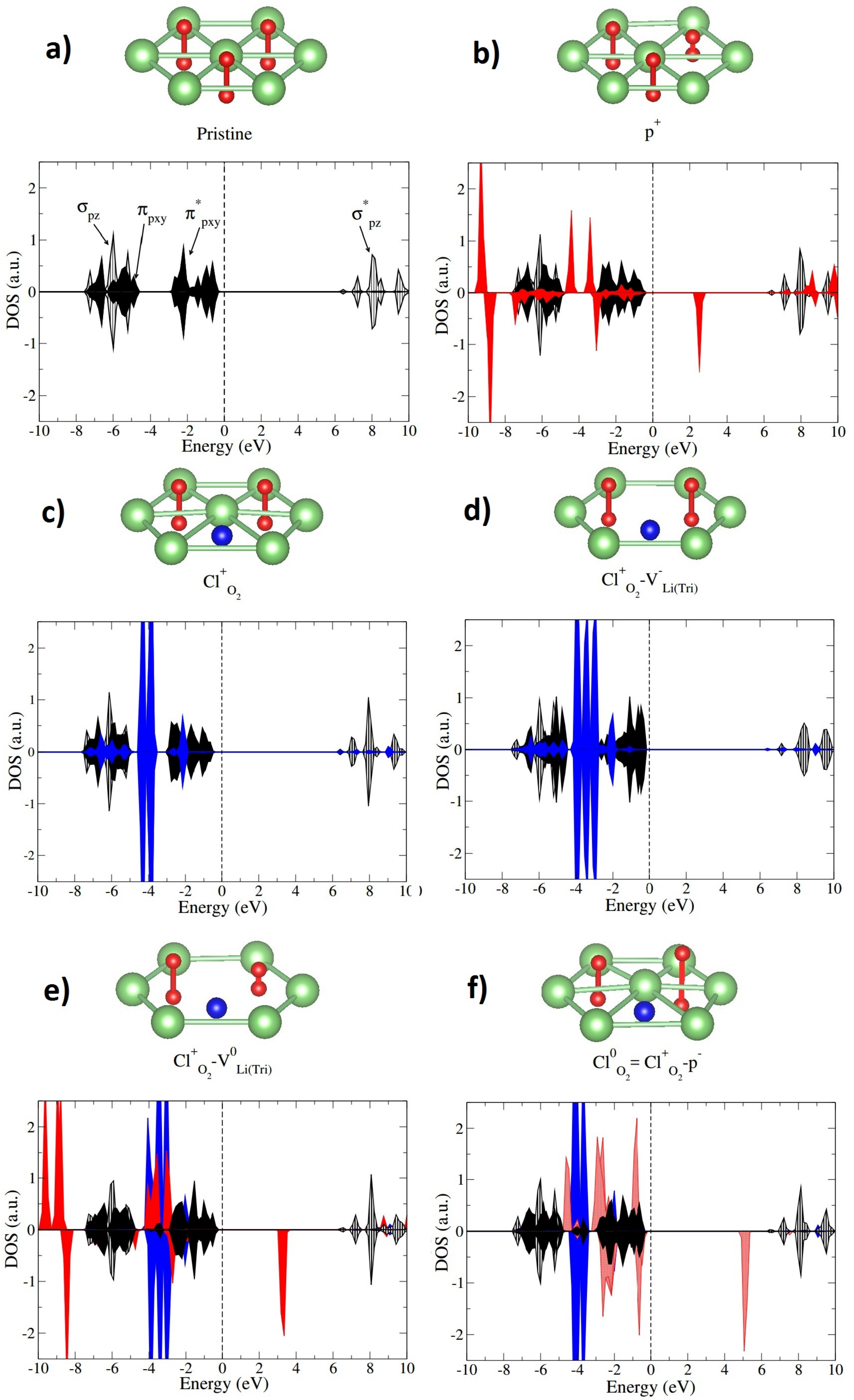}
        \caption{(Color online) Projected DOS onto the p-orbitals of the oxygen dimers in \ce{Li2O2} for the following cases: a) stoichiometric, b) hole polaron $p^+$, c) Cl$^+_{O_2}$, d) Cl$^+_{O_2}$-V$_{Li(Tri)}^-$, e) Cl$^+_{O_2}$-V$_{Li(Tri)}^0$ and f) Cl$^{0}_{O_2}$. Solid (dashes) black curves represents the p$_{xy}$(p$_z$)
orbitals for the O$_2$$^{-2}$ dimers. Solid (dotted) red curves stands for the hole (electron) polaron at the O$_2$$^{-1}$(O$_2$$^{-3}$) dimer, and the blue curves depict the total $p$ states of the Cl ion. On top of each plot, the corresponding structural environment is presented, where green, red and blue spheres represent Li, O, and Cl, respectively.}
        \label{DOS}
\end{figure}

The crystal structure of \ce{Li2O2} is well known, it belongs to the hexagonal \textit{P63/mmc} space group\cite{Cota-2005,Chan-2011}. There are two crystalline sites for Li, with
octahedral and trigonal point groups, respectively, and four oxygens distributed in two dimers per unit cell. Two types of alternated stacking layers are
present, one constituted by the $O_2$-dimers coordinating the trigonal Li site, and the other formed only by the octahedral Li ions.
In Fig. \ref{DOS} a, the calculated density of states (DOS) of stoichiometric \ce{Li2O2} projected onto the p$_{x,y}$ and p$_z$ oxygen orbitals, is depicted.
 The electronic states that determine the chemistry of this crystal are the  $\sigma$($\pi$) bonding and antibonding p$_z$(p$_{x,y}$) orbitals corresponding
to the O$_2$$^{-2}$ dimers.  Both bonding and anti-bonding $\pi$-p$_{xy}$ orbitals are occupied in the stoichiometric situation, while only the bonding $\sigma$-p$_z$ orbital is full.
Fig. \ref{DOS} b, shows in solid red the projected DOS onto the $p$ states of the O$_2$$^{-1}$ dimer. These states correspond to the most energetically favorable intrinsic positive defect, the hole polaron $p^+$ that, basically, consists in a positive charge trapped at an oxygen dimer, whose O-O bond length contracts to 1.32 \AA\ in comparison with 1.47 \AA\ for the O$_2$$^{-2}$ dimer\cite{note1}. 
The structure of the defects is squematized on top of the DOS. Besides the localized states appearing within the bandgap there is an expected
increase of the bonding-antibonding splitting due to the larger Coulomb repulsion of the contracted dimer.
In Fig. \ref{DOS} c-f, the DOS corresponding to some of the more energetically favorable defects associated with Cl are shown.
The notation used for the different type of defects is the same as used by Radin \textit{et al}.\cite{Radin-2015}.
The superindex indicates the effective charge $q$ of the defect, and the subindex indicates the substitutional site being replaced. In Fig. \ref{DOS}c, it is observed that for the case of Cl$^+_{O_2}$, the Cl-$p$ band (in blue) is full and there are no polarons, all the dimers (in black) result in a $-2$ valence state, whose DOS is very similar to the stoichiometric case. It is worth to note that according to Bader charge calculations the Cl ion has valence state $-1$, as happen with the rest of the simulated halogen defects (See the Supplementary section for quantitative information regarding the calculated Bader charges).
In Fig. \ref{DOS} d, for the complex Cl$^+_{O_2}$-V$_{Li(Tri)}^-$ the DOS is similar to the previous case (c) because neither Cl nor the Li vacancy, alter the valence $-2$ of the remaining dimers. At variance with these two last cases, the complex Cl$^+_{O_2}$-V$_{Li(Tri)}^0$ and the defect Cl$^{0}_{O_2}$ do show polaron formation (in red). In the first case, Fig. \ref{DOS} e, it is shown that a hole polaron $p^+$ (solid red) develops, while in the neutral Cl defect case (Fig. \ref{DOS} f) an electron polaron is formed at a neighboring dimer (dotted red).

It is interesting to note, that in the undoped \ce{Li2O2}, the calculated energy gain of localization of an extra charge (a hole polaron) against charge delocalization is around 906 meV.
In the presence of the dopant the delocalized configuration for the extra charge is not even stable\cite{note3}.
The oxygen dimers always prefer to absorb or expel charge in the form of polarons, despite the fact that an energy cost due to the structural distortion have to be paid.
Taking this into account together with the different projected DOS of Fig. \ref{DOS} and the Bader analysis presented in the Supplementary information, 
it can be deduced that the ionic binding of \ce{Li2O2} is not modified by the presence of the dopant. 

Overall, regarding the electronic properties of Cl-doped \ce{Li2O2}, we can conclude that Cl defects will at least promote the formation of electron or hole polarons, but metallic states are not expected to be stabilized. In fact, we have also checked that this situation persists for configurations with higher Cl concentrations\cite{note2}, closer to the ones studied by Matsuda \textit{et al}.\cite{Matsuda-2016}, in which both, electron and hole polarons, remain stable at nearest and next-nearest neighbor dimer sites, respectively. Therefore, whether or not Cl doping can enhance electronic conductivity as suggested by Matsuda \textit{et al}.\cite{Matsuda-2016}, will have to be analyzed in terms of the migration of the above described polarons. Before doing that, it is necessary to calculate the energy cost to form all the studied defects through Eq. \ref{energy}.

\begin{figure}
\centering
\includegraphics[height=7cm]{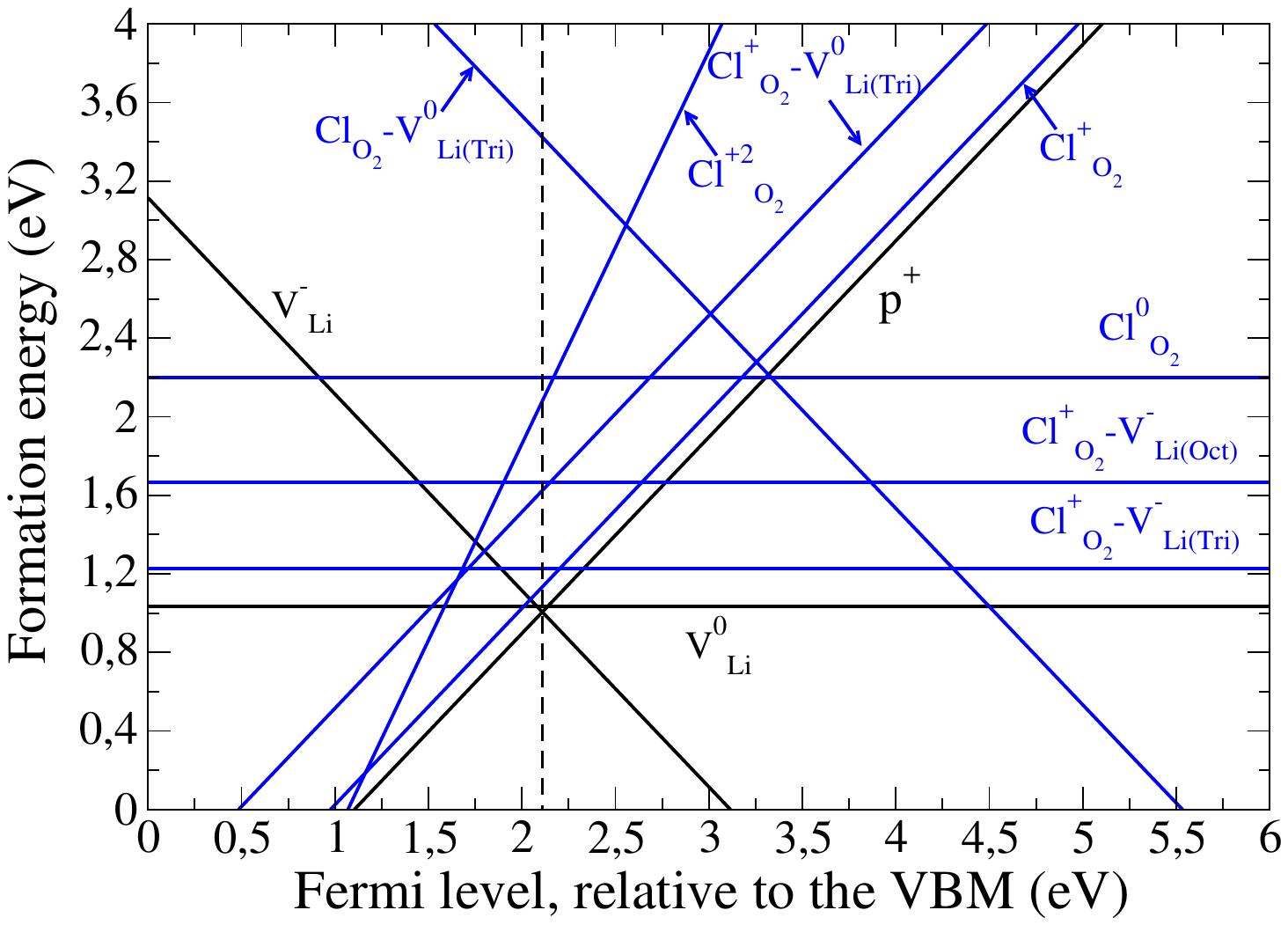}
        \caption{(Color online) Calculated formation energies of all Cl impurities (blue), together with intrinsic \ce{Li2O2} defects (black) as negative Li vacancies, hole polarons and neutral lithium vacancy. Vertical dashed line shows the Fermi level satisfying charge neutrality.}
        \label{Cl-energies}
\end{figure}

Fig. \ref{Cl-energies}  shows the calculated formation energies as a function of the Fermi level of the different defects with Cl, as well as the formation energies of the dominant charged intrinsic defects in \ce{Li2O2}, namely, Li-ion vacancies ($V^{-}_{Li}$), hole polarons ($p^+$), and neutral lithium vacancy ($V_{Li}^0$ = $V_{Li}^-$-$p^+$).

The calculated formation energies and the corresponding equilibrium concentrations  obtained for the Fermi level satisfying electroneutrality condition are listed in Table \ref{energia-Cl}.

\begin{table}
        \centering
        \caption{Calculated formation energies and equilibrium concentrations for intrinsic and Cl-doped \ce{Li2O2} defects using HSE ($\alpha=0.48$). The values are obtained  by setting $E_F$ at charge neutrality.}
  \label{energia-Cl}
  \begin{tabular}{lcc}
    \hline
          Defect  & $E_f(X^q)$ [eV] & $c^0(X^q)$ (cm$^{-3}$)  \\
    \hline
p$^+$ & 0,99 & 2,0x10$^5$ \\
V$_{Li(Tri)}^{-}$ &0,99 &2,0x10$^5$ \\
V$_{Li(Tri)}^{0}$ &1,03 &4,1x10$^4$ \\
Cl$_{O_2}^+$ &1,14 &5,0x10$^2$ \\
Cl$_{O_2}^+$-V$_{Li(Tri)}^{-}$ &1,22 &2,1x10 \\
Cl$_{O_2}^+$-V$_{Li(Tri)}^{0}$ &1,62 &1,2x10$^{-6}$ \\
Cl$_{O_2}^+$-V$_{Li(Oct)}^{-}$ &1,67 &3,1x10$^{-7}$ \\
Cl$_{O_2}^{+2}$ & 2,09 & 7,9x10$^{-15}$\\
Cl$_{O_2}^0$ & 2,20 & 9,7x10$^{-17}$ \\
Cl$_{O_2}^0$-V$_{Li(Oct)}^{-}$ & 3,40 & 1,4x10$^{-37}$ \\
    \hline
  \end{tabular}
\end{table}

First, it should be noted that none of the calculated defects associated with Cl modifies the Fermi level of pristine \ce{Li2O2} if charge neutrality is imposed.
The effect of considering the Cl defects is negligible since the concentration depends exponentially on the formation energy.
The more stable Cl defects are first Cl$^{+}_{O_2}$, followed by Cl$_{O_2}^+$-V$_{Li(Tri)}^{-}$. None of them implies the formation of polarons, and all the dimers remain in the $-2$ valence state.

The neutral Cl defect results in Cl$^{-}$ coupled to an electron polaron formed at a neighboring O$_2$$^{-3}$ dimer. It can be observed from Table \ref{energia-Cl} that this defect is more unstable than Cl$^{+2}_{O_2}$. Thus, it can be concluded that in the presence of the dopant it is easier to create a hole than an electron polaron. It is interesting to remark that a similar situation occurs in undoped \ce{Li2O2}\cite{Radin-2013}. The structural distortion induced by these polarons, on top the distortion induced by the halogen itself, make these kinds of defects very unfavorable.
On the other hand, the complex Cl$_{O_2}^+$-V$_{Li(Tri)}^{-}$ is rather low in energy since it does not imply polaron formation.

Another important remark is that the energy cost of making a Li vacancy in the presence of Cl$^{-}$ is not lower than in pristine \ce{Li2O2}.
Cl$^{-}$ does not promote the Li vacancy formation enhancement, hence our results indicate that it does not behave as a donor dopant as suggested by Gerbig \textit{et al}. \cite{Gerbig-2013} The expected equilibrium concentration of Cl$^{-}$ is two orders of magnitude smaller than the intrinsic defects, so that no dramatic effect can be anticipated in Cl-doped \ce{Li2O2} at equilibrium, a
result that is in line with ionic conductivity measurements  Gerbig \textit{et al}. \cite{Gerbig-2013} If we consider that the system is under the effect of an electrostatic potential as is the case in the measurement by Matsuda \textit{et al}. \cite{Matsuda-2016}, the formation energies of Cl defects increase even more, with the concomitant decrease in the equilibrium concentration. The predicted concentrations of Cl defects are by far smaller than the ones reported by Matsuda \textit{et al}. \cite{Matsuda-2016}, in which the \ce{Li2O2} formation under the operation conditions of the battery might be out of equilibrium.

\begin{figure}
\centering
\includegraphics[height=7cm]{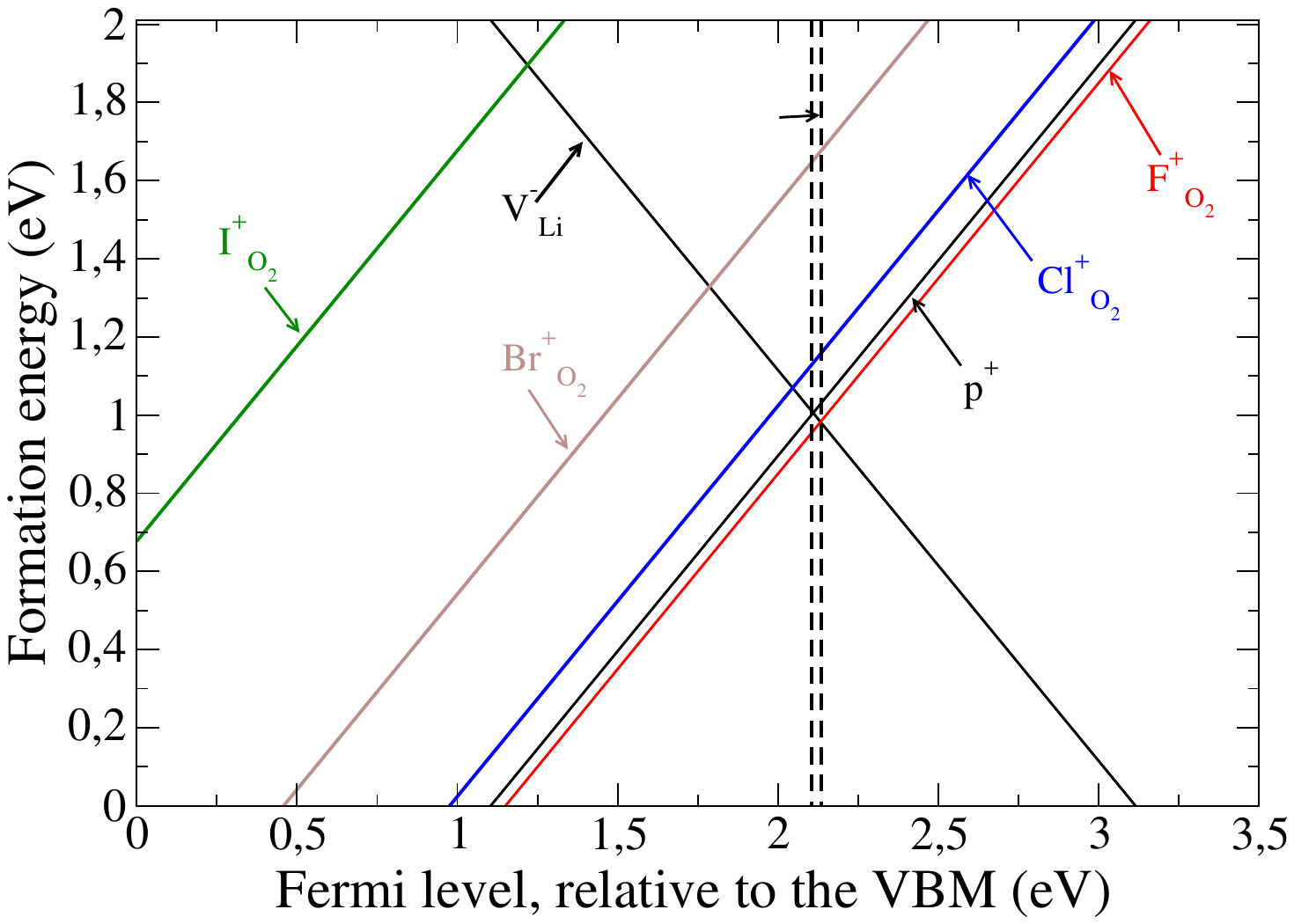}
        \caption{(Color online) Calculated formation energies of all calculated halogen dopants (F, Cl, Br, I) for the corresponding more energetically favorable site, together with intrinsic \ce{Li2O2} defects (black) as negative Li vacancies and hole polarons. Vertical black dashed line shows the Fermi energy satisfying charge neutrality for all cases except for the F-doped system, for which it is depicted with a dotted line, slightly shifted towards larger energies.}
        \label{halogen-energies}
\end{figure}

Fig. \ref{halogen-energies} shows the calculated formation energies of the most stable defect of each halogen, as well as that of the dominant charged intrinsic defects, Li-ion
vacancies (V$_{Li(Tri)}^{-}$) and hole polaron (p$^+$). The obtained values at the corresponding Fermi energy that satisfies charge neutrality in each case, are listed in Table \ref{table-halogens}, along with the strain energy, estimated as follows\cite{Radin-2015}:
\begin{equation}
        E_s= E_{tot}[dist]-E_{tot}[prist] \label{eqn:strain},
\end{equation}
where  E$_{tot}[dist]$ is the total energy of the Li$_2$O$_2$ supercell with no defects, but considering the distorted structural
cell of the doped system, and $E_{tot}[prist]$ is the total energy of pristine \ce{Li2O2}.

The same conclusion as in the Cl-doped system stands for the rest of the halogens regarding the most energetically favorable type of defect. 
As mentioned before, the Bader charge calculation procedure leads to the halogens end up valence state $-1$.
From Fig.\ref{halogen-energies}  and Table \ref{table-halogens} it can be observed that the formation energy of the halogen dopants increases with the size of the ions. In the absence of any other polaronic distortion in these types of defects, this monotonous behaviour can be explained by the increase of the structural strain caused by the bigger dopants. The main conclusion that can be withdrawn from these results is that Cl$^-$ does not show any particular behavior among the halogens. It can be noted that, due to the small size of F, the formation energy of this defect is slightly lower than the intrisic p$^+$. This effect produced a small shift of the Fermi level and, consequently, an increase of the V$_{Li}^{-}$ equilibrium concentration.

\begin{table}
  \centering
        \caption{Formation energies, equilibrium concentrations and strain energies calculated for intrinsic and halogen doped \ce{Li2O2} defects using HSE ($\alpha=0.48$). The values are obtained  by setting $E_F$ at charge neutrality in each case. In parenthesis, the values obtained for the F-doped case Fermi level are provided.}
  \label{table-halogens}
  \begin{tabular}{lccc}
    \hline
          Defect  & $E_f(X^q)$ [eV] & $c^0(X^q)$ (cm$^{-3}$) & E$_s$ (eV) \\
    \hline
          p$^+$ & 0,99 (1,03) & 2,0x10$^5$(4,1x10$^4$) &  1,57\\
          V$_{Li(Tri)}^{-}$ &0,99 (0.95) &2,0x10$^5$(1,0x10$^6$) & 1,23\\
          F$_{O_2}^+$ & 0,96 &6,7x10$^5$ & 0,86\\
          Cl$_{O_2}^+$ & 1,14 &5,0x10$^2$ & 2,00\\
          Br$_{O_2}^+$ & 1,66 &4,7x10$^{-7}$ & 2,56\\
          I$_{O_2}^+$ & 2,80 &7,3x10$^{-27}$ & 3,57\\
    \hline
  \end{tabular}
\end{table}

Thus, we can conclude that F and Cl dopants have rather low formation energies as compared to the \ce{Li2O2} intrinsic defects, so that non negligible concentrations of these defects can be expected. However, from the electronic properties studied in bulk \ce{Li2O2} we do not find any reason why chloride should behave differently from the other halogens as it was suggested by Matsuda \textit{et al}. \cite{Matsuda-2016}

\subsection{Effect of Cl doping on the conductivity of \ce{Li2O2}}

In the previous sections, we have studied the electronic properties of Cl-doped \ce{Li2O2} and concluded that these defects does not induce a change in the electronic conduction mechanism with respect to the pristine \ce{Li2O2} and, consequently, any change in the electronic conductivity have to be analyzed in terms of the migration of the above described polarons.
In particular, we will  calculate the electronic and ionic activation barriers of the more energetically favorable defects, the hole polarons and lithium vacancies,  both in the absence and presence of Cl dopants. We will also analyze the effect of F doping for comparison.

Under these conditions, a diffusion mechanism of polarons and vacancies is expected for the electronic and ionic conductivity, respectively. The
diffusion coefficient is estimated from $D$=$Na^2\nu e^{-\frac{E_a}{k_BT}}$, where $N$ is the number of hopping sites, $a$ is the distance
between hoping sites, $\nu$ is the typical atomic vibration frequency, 
and $E_a$ is the activation barrier. The activation barrier has been estimated differently depending if the defects are lithium vacancies or polarons.

\begin{figure}
\centering
\includegraphics[height=7cm]{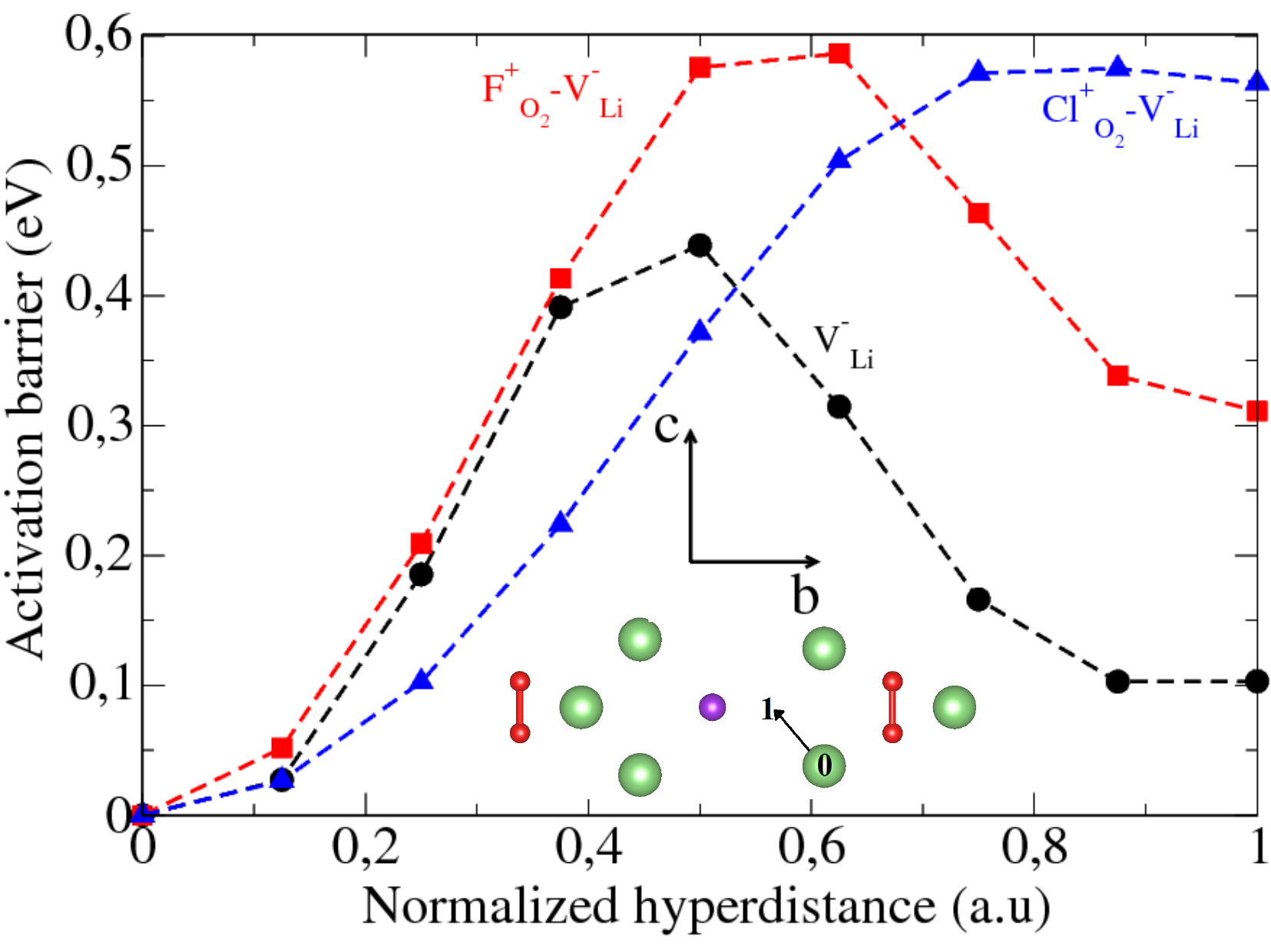}
\includegraphics[height=7cm]{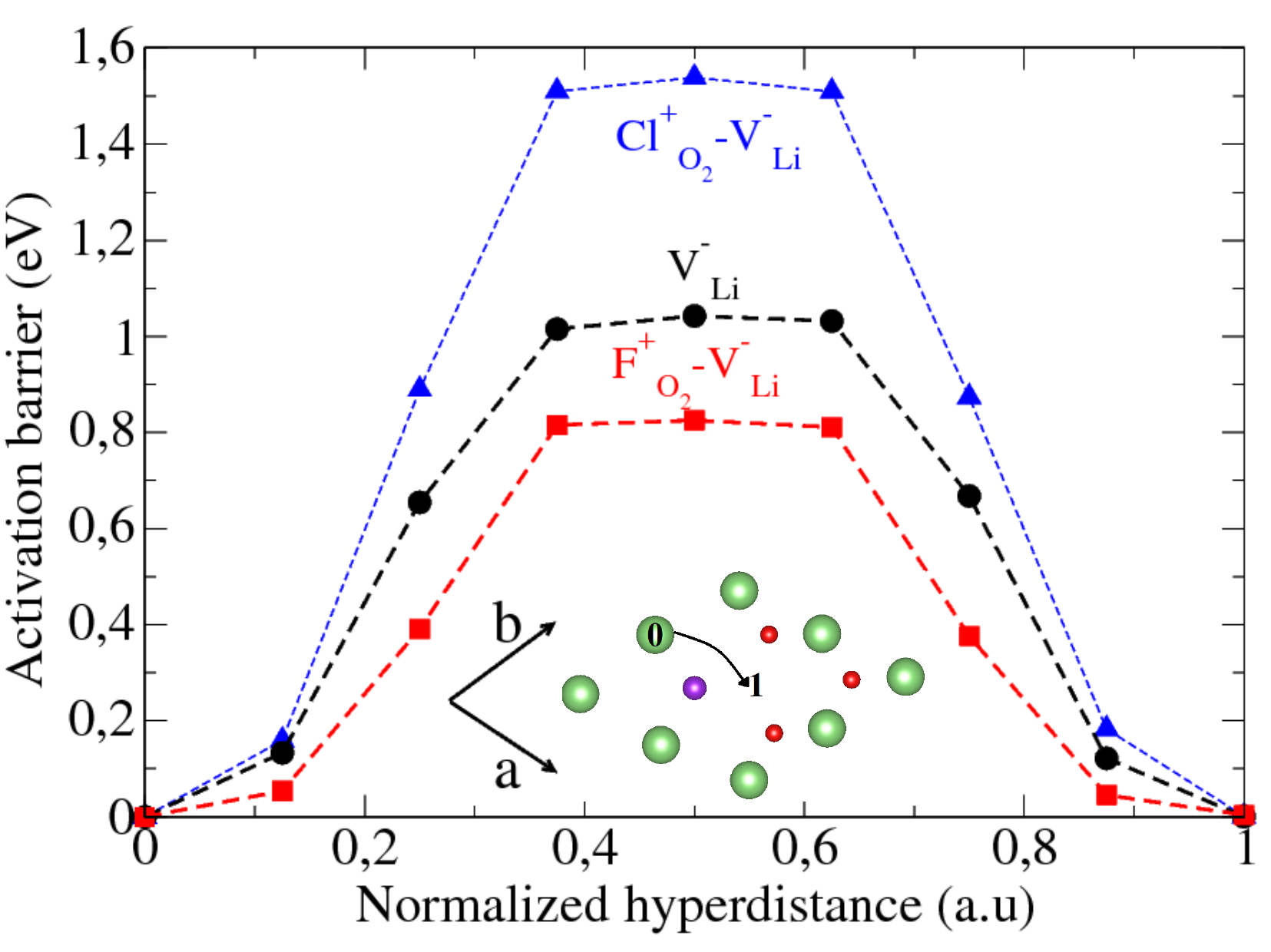}
        \caption{(Color online) Calculated activation energy barrier (E$_a$) for V$_{Li}$ in the presence of a Cl (blue) and F(red) dopants, as well as in the pristine \ce{Li2O2}(black). Upper panel: Interlayer E$_a$. In the abscissa, 0 and 1 correspond to trigonal sites while 0.5 to an octahedral site in between the other two. Bottom panel: Intralayer E$_a$ for the trigonal V$_{Li}$. In this case, 0 and 1 represent two neighboring trigonal sites. The calculated CI-NEB paths are schematized by the dashed arrows in the corresponding insets.}
        \label{barriers}
\end{figure}

In the first case, the climbed-image udged-elastic-band (CI-NEB) method has been used to build the different evaluated diffusion paths. Due to the high computational demand of the CI-NEB method, the vacancy diffusion has been calculated using  PBE as a functional. However, considering  several independent intermediate states we have checked that both, PBE and HSE, give very similar results, confirming that the activation barriers of lithium vacancies are not sensitive to the functional used to calculate it\cite{Radin-2013}. More precisely, we obtain E$_a$=1.33 eV and E$_a$=1.36 eV using PBE and HSE, respectively.
In Fig. \ref{barriers} we show the calculated migration barriers for the inter and intra layer hopping of lithium vacancies. In black we depict the results for undoped \ce{Li2O2} and, in order to assess the role of Cl doping, we study how the activation barrier of V$_{Li}$ is locally affected when passing close to the Cl site (shown in blue). We also show E$_a$ values for the F doping case (in red). In the upper panel of Fig. \ref{barriers}, we plot the interlayer E$_a$. In the inset, we indicate with a black arrow the CI-NEB path in each direction, where 0 and 1 stand for the trigonal and octahedral lithium sites, respectively.
For the interlayer hopping E$_a$$\sim$ 0.45 eV in the undoped case, but this value can locally increase up to 0.57 eV when the vacancy passes close to a Cl site. A similar increment  of E$_a$ is obtained with F doping in this diffusion path.

In the case of intralayer hopping, the calculated E$_a$ are shown at the bottom panel of Fig. \ref{barriers}. The activation barrier of the undoped \ce{Li2O2} is 1.04 eV, substantially larger than for the interlayer diffusion, and this value increases considerably (up to $\sim$ 1.5 eV) when the vacancy passes nearby the Cl site, while it decreases to 0.82 eV when the dopant is F. The decrease of E$_a$ in the last case is due to the smaller size of the F ion.

In summary, regarding the ionic conduction, we have obtained very different values for the inter and intra layer diffusion in the undoped \ce{Li2O2} in good agreement with previous calculations \cite{Radin-2013}, and we conclude that the Cl doping will not promote the formation of lithium vacancies as described in previous sections and it will not enhance their diffusion.

In order to estimate the activation barrier of the hole polarons, we consider the intermediate step of the diffusion path (schematized in Fig. S1 of the supplementary material) and perform the calculation using HSE ($\alpha$=0.48). For the undoped \ce{Li2O2}, we get E$_a$= 0.44 eV and 0.68 eV for the intra and inter layer diffusion, respectively, in good agreement with previous calculations\cite{Radin-2013, DFT+U}. At variance with the ionic conduction, the interlayer path is preferred in this case. The intralayer barrier increases to 0.57 eV when the hole polaron passes close to a Cl site, while the interlayer one is not altered by the presence of the dopant. On the other hand,  due to the smaller size, the intra and interlayer hopping decrease nearby the F site to 0.41 eV and 0.64 eV, respectively.

Thus, our conclusions for the electronic conduction in \ce{Li2O2} in the presence of Cl dopants is similar to that of the ionic one. Cl dopants are not expected to modify neither, the electronic conduction mechanism, nor the concentration of intrinsic \ce{Li2O2} hole polarons. Moreover, the results indicate that chloride will not favor the polaron mobility.
In fact, it is shown that F doping could slightly improve the electronic conductivity.

Thus, our results do not support the conclusions by  Matsuda \textit{et al}. \cite{Matsuda-2016} that \ce{Li2O2} enhanced conductivity by Cl doping is responsible for the high LAB capacity observed. Indeed, some caution should be taken in the interpretation of the conductivity changes in \ce{Li2O2}  upon doping (see Supporting Information for details), and we suggest that further experimental studies should be performed before accepting that Cl doping affect so dramatically the conductivity of LABs discharge products.

\section{Conclusions}

By resorting to first-principle calculations we studied in detail the effect of several halogen dopants on the bulk electronic and transport properties of \ce{Li2O2}. Our results describe the effects of these dopants, in particular Cl, well inside the nano particles or toroids formed during the discharge  of the LAB.
In that sense, all the energetic has been done assuming charge neutrality.

The rather low calculated formation energies of F and Cl defects, as compared to the intrinsic defects of \ce{Li2O2} indicate that non-negligible concentrations of these dopants can be expected, as it was reported for the case of Cl-doped \ce{Li2O2}\cite{Gerbig-2013,Matsuda-2016}.
The other halogen dopants, Br and I, induce larger structural distortions, making them very energetically unfavorable.
Our results do not indicate any special behavior of Cl among the halogen ions. The formation energies exhibit a reasonable dependence with  the dopant ionic size.

The most stable defect, associated with Cl as a dopant, is the positively charged one replacing an oxygen dimer O$^{-2}_2$, in agreement with previous calculations \cite{Radin-2013}.
The same conclusion holds for the rest of the studied halogens F, Br, and I.
In general, for these defects there is no extra electron or hole polaron formation, i.e. the oxygen dimers are all in a -2 charge state.


Our results for Cl-doped \ce{Li2O2} do not support any of the mechanisms for enhancing conductivity, namely, the excess of charge promoting metallic states that preempt the polaron formation\cite{Vladimir-2013}, the shift of the Fermi level modifying the charge balance of the system \cite{Radin-2015}, or the creation of extra charge carriers beside the intrinsic ones of \ce{Li2O2}. 
In all the cases (doped and undoped) where there is extra charge, the system prefers to pay the cost of the structural distortion rather to let this extra charge gets delocalized. Moreover, this effect is more a consequence of a \ce{Li2O2} characteristic than the nature of the particular dopant, since the ionic binding of the crystal and the capability of the \ce{O2} dimer to change its valence remains in the presence of the dopant.

We found that Cl doping does not promote the formation of extra lithium vacancies and polarons, and neither it induces a change in the conduction mechanism, even at the high concentrations experimentally achieved \cite{Matsuda-2016}. The calculated activation barriers of both hole polarons and lithium vacancies
increase in the Cl-doped system, indicating that no improvement of the bulk conductivity through \ce{Li2O2} can be expected in this case.

It should be emphasized that these conclusions could not be straightforward assumed to be valid in the case of thin films or small nanoparticles, where surfaces conductivity or other mechanism could play a relevant role. Further studies are desired regarding the relation between the  morphology and the conductivity  of doped \ce{Li2O2}, a still open question even for the undoped case, that could explain the reported increase of the capacity of LABs when Cl is added as a dopant. Regarding this experimental finding, one should be cautious about its relation with an increase in the bulk conductivity of the Cl-doped \ce{Li2O2}.


\section*{Acknowledgements}
The authors thank financial support from ANPCyT (PICT 2015 0869, PICT 2014 1555, PICTE 2014 134) and CONICET (PIP 2015 0364 GI). The authors kindly thank Maxwel Radin for useful information regarding the calculation of the chemical potentials. VLV, MAB, and HRC are members of CONICET. HACP thanks a fellowship from ANPCyT.



\balance


\bibliography{lab,notes} 

\begin{mcitethebibliography}{37}
\bibitem[note1()]{note1}
With HSE and $\alpha$=0.48 the O-O bond legth is 1,47 \AA, with $\alpha$=0.25
  it results 1.50 \AA\ and with DFT-PBE 1.54 \AA .\relax
\mciteBstWouldAddEndPunctfalse
\mciteSetBstMidEndSepPunct{\mcitedefaultmidpunct}
{}{\mcitedefaultseppunct}\relax
\EndOfBibitem
\bibitem[note1()]{note3}
The reason why the delocalized configuration is not stable in the presence of
  the dopant, is because the oxygen dimers close to it are further away from
  each other (in average) so that is not possible to form metallic bands.\relax
\mciteBstWouldAddEndPunctfalse
\mciteSetBstMidEndSepPunct{\mcitedefaultmidpunct}
{}{\mcitedefaultseppunct}\relax
\EndOfBibitem
\end{mcitethebibliography}
\bibliographystyle{rsc} 

\end{document}